\documentclass[12pt]{article}
\newcommand{\beq}{\begin{equation}}
\newcommand{\eeq}{\end{equation}}
\newcommand{\bea}{\begin{eqnarray}}
\newcommand{\eea}{\end{eqnarray}}

\newcommand{\gsim}{\lower.7ex\hbox{$
\;\stackrel{\textstyle>}{\sim}\;$}}
\newcommand{\lsim}{\lower.7ex\hbox{$
\;\stackrel{\textstyle<}{\sim}\;$}}

\newcommand{\eod}{\end{document}}

\begin{document}
\thispagestyle{empty}
\vspace*{-22mm}

\begin{flushright}
UND-HEP-12-BIG\hspace*{.08em}11\\


\end{flushright}
\vspace*{1.3mm}

\begin{center}
{\Large {\bf Probing CP Asymmetries in Charm Baryons Decays}}

\vspace*{10mm}

{ I.I.~Bigi$^a$}\\
\vspace{7mm}
$^a$  {\sl Department of Physics, University of Notre Dame du Lac, Notre Dame, IN 46556, USA}\\

{\sl email addresses: ibigi@nd.edu} \\

\vspace*{10mm}

{\bf Abstract}
\vspace*{-1.5mm}
\\
\end{center}

With the first evidence of direct CP violation in $D^0 \to K^+K^-/\pi^+\pi^-$ 
there is a strong reason to probe CP asymmetries in charm baryons in three-body 
final states $\Lambda_c \to p\pi^+\pi^-/pK^+K^-$ \& 
$\Lambda_c \to pK^+\pi^-$ with $\Lambda_c \to pK^-\pi^+$ to calibrate them. 
Analyzing the Dalitz plots carefully allows to find the `existence' of New Dynamics (ND) and its `features'. One can test different methods to probe CP asymmetries. Final data from CDF and D0 experiments have advantage by comparing $\Lambda_c $ with 
$\bar \Lambda_c $ directly from $p \bar p $ collisions. LHCb will need more data with 
$\bar \Lambda_c$ transitions.

\vspace{3mm}

\hrule

\tableofcontents
\vspace{5mm}

\hrule\vspace{5mm}

\section{CP Asymmetries in Baryons Decays} 

So far, no evidence has been found about CP violation in baryons -- 
except `our' existence. Now we have seen the first evidence for (direct) CP violation in the dynamics 
of up-type quarks, namely asymmetry in $D^0 \to K^+K^-/\pi^+\pi^-$ \cite{LHCb1,CDF1} that could be 
a sign of the impact of New Dynamics (ND) beyond CKM forces. It should `whet' our appetite for 
finding CP asymmetries in charm baryons in singly and doubly Cabibbo suppressed decays 
(SCS and DCS) like $\Lambda_c^+ \to p \pi^+\pi^-$ vs. 
$\bar \Lambda_c^- \to \bar p \pi^-\pi^+$ or $\Lambda_c^+ \to p K^+\pi^-$ 
vs. $\bar \Lambda_c^- \to \bar p K^-\pi^+$. Cabibbo favoured $\Lambda_c^+ \to p K^-\pi^+$ can 
be used to calibrate branching ratios and also CP asymmetries, since ND have hardly a chance to 
produce measurable CP violation there. 

\subsection{Parameterization CKM Matrix through ${\cal O}(\lambda ^6)$}

Yet averaging data on semileptonic $B_d$, $B_u$ decays PDG states 
$|V_{ub}/V_{cb}|$ around 0.08. It means 
one has to use a parametrization through $O(\lambda^6)$ and with other quantities 
of true order of unity. One has been found specifically in Ref.\cite{AHN} with $\lambda$, $f \sim 0.75$, $\bar h \sim 1.35$ and $\delta_{\rm QM} \sim 90^o$:  
\begin{eqnarray} 
V _{\rm CKM} = 
\left(\footnotesize
\begin{array}{ccc}
 1 - \frac{\lambda ^2}{2} - \frac{\lambda ^4}{8} - \frac{\lambda ^6}{16}, & \lambda , & 
 \bar h\lambda ^4 e^{-i\delta_{\rm QM}} , \\
 &&\\
 - \lambda + \frac{\lambda ^5}{2} f^2,  & 
 1 - \frac{\lambda ^2}{2}- \frac{\lambda ^4}{8}(1+ 4f^2) 
 -f \bar h \lambda^5e^{i\delta_{\rm QM}}  &
   f \lambda ^2 +  \bar h\lambda ^3 e^{-i\delta_{\rm QM}}   \\
    & +\frac{\lambda^6}{16}(4f^2 - 4 \bar h^2 -1  ) ,& -  \frac{\lambda ^5}{2} \bar h e^{-i\delta_{\rm QM}}, \\
    &&\\
 f \lambda ^3 ,&  
 -f \lambda ^2 -  \bar h\lambda ^3 e^{i\delta_{\rm QM}}  & 
 1 - \frac{\lambda ^4}{2} f^2 -f \bar h\lambda ^5 e^{-i\delta_{\rm QM}}  \\
 & +  \frac{\lambda ^4}{2} f + \frac{\lambda ^6}{8} f  ,
  &  -  \frac{\lambda ^6}{2}\bar h^2 , \\
\end{array}
\right)
+ {\cal O}(\lambda ^7)
\end{eqnarray}
The input for this parametrization is close to the data given by HFAG, in particular for $|V_{ub}/V_{cb}|$ 
averaged over values from $B \to l \nu \pi$ and $B\to l \nu X_c$. The central value is actually close to 
$|V_{ub}|_{\rm excl.}$ rather than the larger $|V_{ub}|_{\rm incl.}$. 

The pattern in flavour dynamics is less obvious and more subtle for CP violation: 
\bea
{\rm Class\; I.1:}&&V_{ud}V^*_{us} \; \; \;  [{\cal O}(\lambda )] + V_{cd}V^*_{cs} \;  \; \;  [{\cal O}(\lambda )] + 
 V_{td}V^*_{ts} \; \; \; [{\cal O}(\lambda ^{5\& 6} )] = 0   \\ 
{\rm Class\; I.2:}&& V^*_{ud}V_{cd} \; \; \;  [{\cal O}(\lambda )] + V^*_{us}V_{cs} \; \; \;  [{\cal O}(\lambda )] + 
V^*_{ub}V^*_{cb} \; \; \; [{\cal O}(\lambda ^{6 \& 7} )] = 0    \\
{\rm Class\; II.1:}&& V_{us}V^*_{ub} \; \; \;  [{\cal O}(\lambda ^5)] + V_{cs}V^*_{cb} \;  \; \;  [{\cal O}(\lambda ^{2 \& 3} )] + 
V_{ts}V^*_{tb} \; \; \; [{\cal O}(\lambda ^2  )] = 0   \\ 
{\rm Class\; II.2:}&& V^*_{cd}V_{td} \; \; \;  [{\cal O}(\lambda ^4 )] + V^*_{cs}V_{ts} \; \; \;  [{\cal O}(\lambda ^{2\& 3})] + 
V^*_{cb}V^*_{tb} \; \; \; [{\cal O}(\lambda ^{2 \& 3} )] = 0  \\
{\rm Class\; III.1:}&& V_{ud}V^*_{ub} \; \; \;  [{\cal O}(\lambda ^4)] + V_{cd}V^*_{cb} \;  \; \;  [{\cal O}(\lambda ^{3\& 4} )] + 
V_{td}V^*_{tb} \; \; \; [{\cal O}(\lambda ^3  )] = 0   \\ 
{\rm Class\; III.2:}&& V^*_{ud}V_{td} \; \; \;  [{\cal O}(\lambda ^3 )] + V^*_{us}V_{ts} \; \; \;  [{\cal O}(\lambda ^{3\& 4})] + 
V^*_{ub}V^*_{tb} \; \; \; [{\cal O}(\lambda ^4 )] = 0
\eea  
\begin{itemize}
\item
As is well known, direct CP violation is allowed in {\em singly} Cabibbo suppressed $D$ decays in the SM -- 
the question is on which level. Using the modified QM parametrization from Ref.\cite{AHN} one gets 
$\Delta A_{CP} \propto  f\bar h \lambda^5 \times \sin \delta_{\rm QM} \simeq (0.5 - 0.6)\times 10^{-3}$
for $\delta_{\rm QM} = 75^o - 120^o$.   

`Penguin' diagrams show such mixing can happen and produce sizable $SU(3)_{\rm flav}$ violation first suggested in \cite{SANDA1} and very recently with detail about U-spin in \cite{BRODMARCH}. However Penguin diagrams will contribute little to charm baryons for three-body final states.

\item 
SM amplitudes 
for DCS transitions are of ${\cal O}(\lambda ^2)$; on the quark level there is only one 
amplitude with a weak phase at most of 
${\cal O}(\lambda ^5) \sim (0.5 - 0.6) \times 10^{-3}$. For finding the impact of ND in a realistic  
situation we can ignore SM weak phase.

\end{itemize}

\section{CP Asymmetries in Charm Baryons}

Charm baryons might to turn out the `Poor Princesses' for establish CP violation in baryons decays 
and even to show the impact of ND there. That is based on several reasons: 
\begin{itemize}
\item 
The `background' from SM for CP asymmetries is very small or even zero. 
\item 
The ratios of three-body final states are larger than for two-body ones. 
\item 
In particular one can directly measure them in $p \bar p$ collisions now at FNAL and in the 
future at FAIR by PANDA experiment; likewise in the future in $e^+e^-$ annihilations at Super-Belle and SuperB experiments.  
\item 
Three-body final states produce many CP odd observables unlike in two-body final states. 
{\em Relative} asymmetries inside the Dalitz plots do not depend on the {\em productions} of the 
decaying baryons. 
\item 
One gets final states of all charged hadrons in $\Lambda_c^+ \to p \pi^+\pi^-/pK^+K^-$ and 
$\Lambda_c^+ \to p K^+\pi^-$, where one can probe for CP asymmetries and calibrate them 
with $\Lambda_c^+ \to p K^-\pi^+$.

\end{itemize}

While analyzing the two-dimensional  Dalitz plots needs larger amounts of data and experimental work, 
but they also deliver `profits' \& `prizes', namely about the {\em existence} of ND and its 
{\em features}. One can use model {\em in}dependent analyses: 
\begin{itemize}
\item 
 `Miranda I' \cite{MIR1} uses `significance' 
 \beq 
 S_{CP}(i) \equiv \frac{N(i) - \bar N(i)}{\sqrt{N(i) + \bar N(i)}}
 \eeq
 for the bin $i$ rather than studies the 
`fractional' asymmetries $\Delta (i) = [N(i) - \bar N(i)]/[N(i) + \bar N(i)] $; 
its strength is {\em localizing} CP violation.

\item 
The refined `Miranda II' \cite{MIR2} follows where each bin has $N = N^+ + N^-$ events with 
$N^+$ and $N^-$ being the numbers of $p$ and $\bar p$ inside. $N^+$ follows a 
binomial distribution with `expected value' and `variance' given by 
\beq 
E [N^+] = N P \; \; , \; \; V[N^+] = N P (1-P)  
\eeq  
If $N$ is large enough, one has 
\beq
A^{\rm bin}_{CP} = \frac{N^+ - N^-}{N} 
\eeq
with 
\bea 
\mu &=& E [ A^{\rm bin}_{CP} ] = 2P - 1 \\
\sigma ^2 &=& V[A^{\rm bin}_{CP}] = \frac{4P(1-P)}{N} .
\eea
It needs more working, yet it helps significantly the {em features} of CP odd forces. 

\item  
One can also use the method suggested in Ref.\cite{WILL}. 

\end{itemize}  

Those analyses should not be the final steps for getting lessons about ND.  
Final state interactions (by strong forces) cannot be calculated from first principles now. However 
one can relate them to low-energy $\pi K/2\pi/K^+K^-$ and $p\pi/pK$ scattering -- 
with some {\em non}-trivial 
theoretical tools using `dispersion relations' \cite{KUBIS}. 

\subsection{DCS $\Lambda_c^+ \to p K^+\pi^-$ vs. $\bar \Lambda_c^- \to \bar p K^-\pi^+$}

As mentioned above, SM produces only one quark amplitude for DCS transitions; therefore SM cannot produce CP asymmetry. Furthermore the size of SM amplitudes are very much suppressed 
to give more sensitivity to the impact of ND.

In particular, one can analyze 
$\Lambda_c^+ \to p K^+\pi^-$ vs. $\bar \Lambda_c^- \to \bar p K^-\pi^+$  about CP violation and compare with CF $\Lambda_c^+ \to p K^-\pi^+$ \& $\bar \Lambda_c^- \to \bar p K^+\pi^-$
to learn about the impact of final states interactions. Of course, one has to differentiate $K^+\pi^-$ from 
$K^-\pi^+$ in the $\Lambda^+_c$ decays despite the huge difference in their branching ratios. 
These DCS decays have not been found yet. On the other hand one can hope for significant CP violation in the DCS transitions.  

The $\Lambda_c$ final states include $pK^*$, $p\kappa$, $N^*K$, $\Delta ^{(*)}K$, 
$\Lambda^*\pi$ etc.  -- i.e., numerous states to give us lessons about the existence of ND and its  
features due to several reasons: 
\begin{itemize}
\item 
In the SM there is only one quark amplitude -- therefore no CP asymmetry. 
\item 
The SM amplitudes are significantly suppressed by order of tg $\theta_C^2$  
and transitions by order of tg $\theta_C^4$; thus the sensitivity for ND's impact is 
larger. 
\item 
One can ignore CKM phases in `reality'. 
\item
The `local' CP asymmetries are usually larger than the `global' asymmetry -- i.e., 
averaged over the local ones. 

\end{itemize}
These are qualitative and at most a semi-quantitatively comments. Quantitative theoretical works 
will happen based on dispersion relations, but they will take more efforts and time 
(in particular for $pK$ \& $p\pi$ states). 

\subsection{SCS $\Lambda_c^+ \to p \pi^+\pi^-/pK^+K^-$}

While the SCS $\Lambda_c^+ \to p \pi^+\pi^-/pK^+K^-$ have been seen, but not established -- 
namely branching ratios of $(3.5 \pm 2.0) \times 10^{-3}/(7.7 \pm 3.5) \times 10^{-3}$.   
Existing and/or future data should be able to analyze them in details from CDF/D0 and 
LHCb experiments. The `landscapes' of their Dalitz plots give many avenues to find CP asymmetries and ND as probed in DCS decays; however there are complications: 
\begin{itemize} 
\item 
There are `backgrounds' from SM. 
\item 
SM amplitudes are less suppressed and therefore give less sensitivity for ND amplitudes. 

\end{itemize}

For these SCS decays CKM dynamics produce a weak phase for $c \to s \bar s u$ of order 
$\lambda^5$, but not for $c \to d \bar d u$. However, intermediated states of $s\bar s$ mix 
with $d \bar d$; therefore very small CP asymmetries are likely to appear {\em both} in 
$\Lambda_c^+ \to p \pi^+\pi^-$ with $p\rho^0$, $p \sigma$ etc. and 
$\Lambda_c^+ \to p K^+K^-$ with $p\phi$, $\Lambda^*K$ etc. The final states of $p\pi^+\pi^-$ and $pK^+K^-$ are complex, but not as much as for DCS transitions and with a SM `background'. 
As before, the `local' CP asymmetries are usually larger than the `global' asymmetries in 
three-body final states, and quantitatively theoretical analyses can and should be worked based on 
dispersion relations.

\section{Summary}
   
No CP violation has been found in strange, beauty and charm {\em baryons} dynamics. 
Charm baryons 
and in particular $\Lambda_c$ decays give good opportunity to establish CP asymmetries 
in three-body final states with charged hadrons: 
\begin{itemize} 
\item 
For DCS $\Lambda_c^+ \to p K^+\pi^-$ vs. $\bar \Lambda_c^- \to \bar p K^-\pi^+$ there is 
{\em no} `background' from SM. Furthermore the impact of ND gives us lessons of the features 
of ND in these three-body states. 
\item 
SCS $\Lambda_c^+ \to p \pi^+\pi^-/pK^+K^-$ transitions have a chance to find CP violation in 
baryons decays; it has to deal with non-zero `background' for SM. 

\item 
Existing CDF/D0 data, future LHCb data and even more future data from PANDA, Super-Belle 
and SuperB will tell us significant lessons. 

\item 
More insights about the underlying ND will give us based on 
non-trivial theoretical analyses of dispersion relations.  

\end{itemize}

\vspace{0.5cm}

{\bf Acknowledgments:} This work was supported by the NSF under the grant number PHY-0807959. 

\vspace{4mm}


\end{document}